# Prospects for studies of high-energy solar cosmic rays with ATLAS

S. N. Karpov, Z. M. Karpova, V. A. Bednyakov

*Joint Institute for Nuclear Research (JINR), Dubna, Russia*

The ATLAS detector is intended to verify the standard model and to search for new physics at the Large Hadron Collider (LHC, CERN). In addition to this primary goal, it also allows detection of muons of cosmic rays. On the other hand, unusual bursts of the muon intensity, which correlate with powerful solar flares were recorded and investigated earlier at the Baksan Underground Scintillation Telescope (BUST, INR, Russia) in period from 1981 to 2006 ($\approx$2.5 solar cycles). The nature of the muon bursts and their probable relation to the solar cosmic rays is still not quite clear. ATLAS has an excellent muon system allowing search for similar muon bursts. Within the next few years, when the LHC and ATLAS should start to operate, an increase in the solar activity is expected in the new 24th cycle. It increases the probability of finding the muon bursts from powerful flares. Hence ATLAS has a good opportunity to verify the relation of muon bursts to the solar cosmic rays.

PACS: 96.50.S-, 96.50.Vg, 96.60.-j, 96.60.qe – cosmic rays, energetic particles, solar physics, flares.

## 1 Introduction

As is well-known, the Sun continuously emits huge flows of plasma named a solar wind in addition to light and heat. There are also high-speed streams of plasma from open magnetic areas of the Sun and coronal mass ejections during powerful flares. The speed of such streams is several times higher than speed of usual solar wind. In all these cases the solar plasma consists basically of protons and electrons with rather low energy (in the ranges of keV and MeV). Such particles cannot reach the surface of the Earth. Strong enough magnetic field of the Earth (magnetosphere) alters motion of the solar plasma. The solar wind flows round the Earth along the external border of the magnetosphere. Some part of the solar wind particles can be captured into radiation belts of the Earth. They are part of the magnetosphere located noticeably higher than dense layers of the atmosphere.

However, during powerful solar flares some part of the solar plasma particles can be accelerated to a sufficiently high energy. Streams of such particles are named solar cosmic rays (SCR). Under certain conditions protons can be accelerated on the Sun and in the interplanetary space to units and tens of GeV. Such particles are capable of penetrating the magnetosphere, atmosphere and reaching the surface of the Earth. In this case the streams of SCR from solar flares become visible as rather short-term increases (from tens of minutes to tens of hours) in intensity of cosmic rays (in a counting rate of the detector). The SCR protons as well as galactic cosmic rays (GCR) interact with nuclei at passing through the atmosphere. As a result, several components of secondary cosmic rays will be registered on the terrestrial surface: the electron-photon component, muons and the hadron-nucleus component. The intensity increases caused by the SCR, which reached the terrestrial surface are usually registered and investigated with the help of a world-wide network of Neutron Monitors (NM). They register secondary neutrons resulting from interactions of primary protons in the atmosphere. It is not in every solar flare that protons are accelerated to energy more than 1 GeV, which is necessary for reaching the terrestrial surface. Moreover, the SCR particles usually move from the Sun to the Earth along lines of the interplanetary magnetic field. If the flare area has unsuccessful location on solar disk the stream of SCR can pass by the Earth. Therefore, cases where neutron monitors register the intensity increases caused by the SCR flux are treated as a separate class of solar events. Such events are usually called Ground Level Enhancements or the GLE events. In fig. 1 (bottom panel) one can see the example of the Ground Level Enhancement of SCR, which was recorded by the world-wide network of neutron monitors on September 29, 1989. The intensity increase which is produced by the SCR is shown in percents concerning the average background of galactic cosmic rays $N_{GCR}$ measured prior to the beginning of flare: $A(\%) = (N - N_{GCR}) / N_{GCR} \cdot 100\%$. The intensity of cosmic rays at the terrestrial surface can be increased during the GLE event in some times (several hundreds of percents) as it can seen in fig. 1. The amplitude and duration of signal strongly decrease with increase of the energy of the SCR particles to which the detectors are sensitive.

Within the active phase of a solar cycle various flares occur practically every day and sometimes up to ten per day. At the same time, the average frequency of the Ground Level Enhancements is only ~1 event per year. These events are usually grouped at the phases of increase and recession of activity of a solar cycle. But their number can only rarely approach ~10 per year in the active phase of the solar cycle. The neutron monitors usually record increases of intensity from SCR with the energy no more than 4-5 GeV. Insufficient statistical accuracy of NM does not allow measuring the SCR flux at higher energy. The NM signals with energy of SCR as high as 15-20 GeV were recorded only in two huge GLE events associated with flares on February 23, 1956 and September 29, 1989 [1]. Therefore, it was long believed, that protons can be accelerated on the Sun to no more than 20 GeV. It quite agrees with some acceleration models in which the SCR spectrum abruptly goes down at high energy. It is the so-called cut-off of the solar spectrum at high energy.

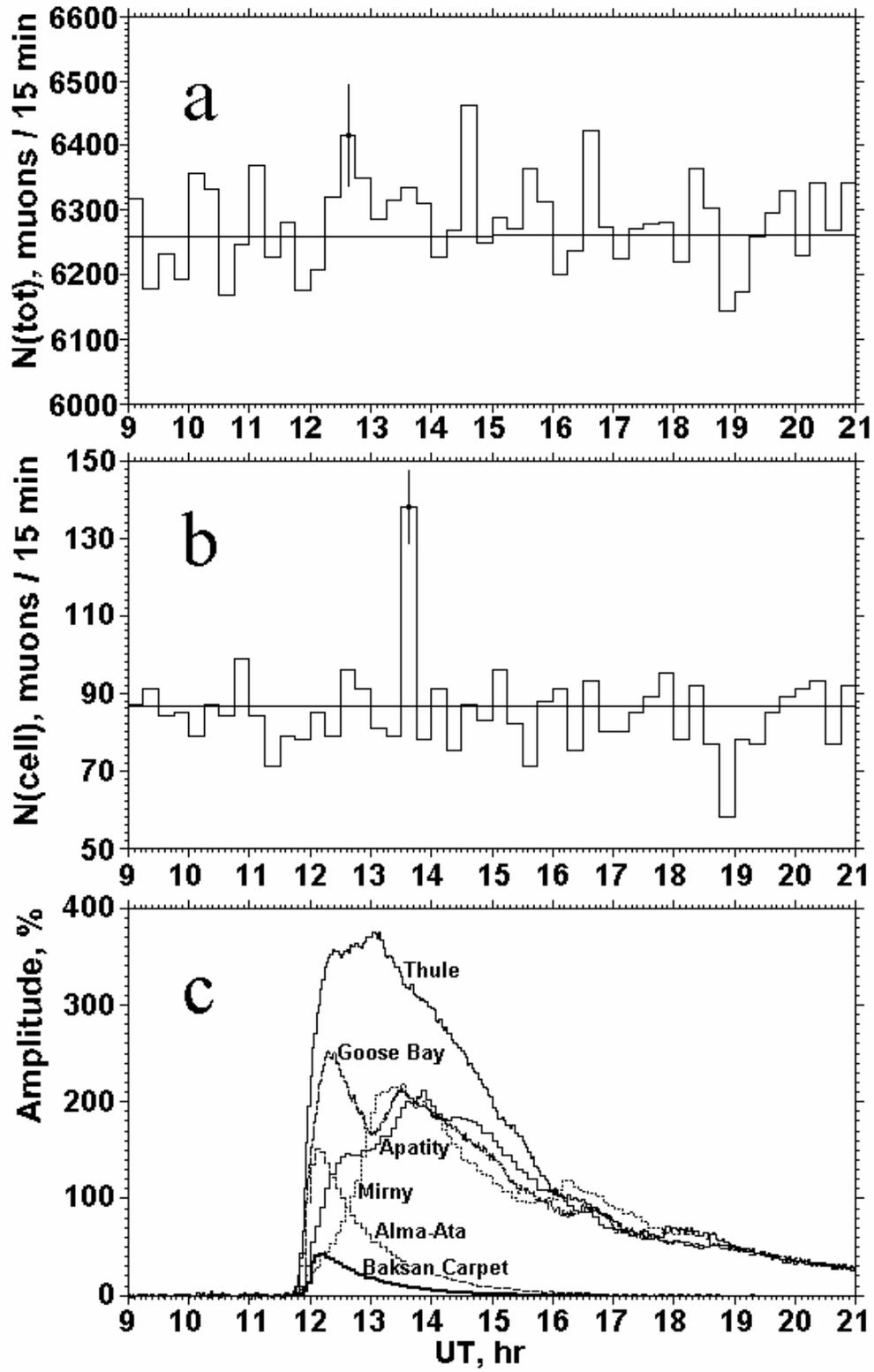

**Fig.1.** Time profiles of the cosmic ray intensity at various detectors on September 29, 1989 [7]. Bottom panel: the SCR intensity increases at several neutron monitors (names of the NM stations are shown near lines); middle panel: burst in muon intensity at the BUST in a narrow solid angle (in angular cell); top panel: the total counting rate of muons at the BUST.

However, during the GLE event on September 29, 1989 the signals of the SCR were recorded not only by neutron monitors. For the first time the increase in the SCR flux was recorded by the Baksan CARPET detector [2, 3] which is intended for registration of Extensive Air Showers (EAS). Moreover, the signals were registered by several underground muon detectors [4, 5], also for the first time. Finally an unusual burst of muon intensity was abserved at the Baksan Underground Scintillation Telescope (BUST) [6]. Additional muons were recorded at the BUST in a narrow enough solid angle – in a cell 10 × 15 degrees in size. This burst of the muon intensity is shown in the middle panel of fig. 1 [7]. The statistically significant signal was not found in the analysis of the total counting rate of muons at the BUST (top panel in fig. 1). Similar muon bursts were also observed at the BUST later during several other GLE events [7, 8]. Statistical significance and some features of muon bursts recorded by the BUST do not allow stating unambiguously that increases of the muon intensity are caused by the solar cosmic rays. However, other possible interpretations meet even greater difficulties.

Registration of the above-mentioned signals raised the question about a possibility to accelerate of protons on the Sun to several hundreds of GeV. Some models of acceleration [9-11] theoretically allow protons with such energy and even up to hundreds of TeV on the Sun [12]. However it remains not clear whether necessary conditions can actually occur in the regions of flares (a configuration of magnetic and electric fields, density and speed of plasma, etc.). Therefore, the question about the upper limit and the form of the energy spectrum of the solar cosmic rays remains open [13].

## 2 Muon bursts at the BUST

Unusual short-term bursts of the muon intensity correlated with the powerful solar flares were registered and investigated [7, 8] at the Baksan Underground Scintillation Telescope during almost 2.5 cycles of solar activity (from middle of 21st to end of 23rd cycles: 1981-2006). Speaking more precisely, the unusual muon bursts were recorded during some events of the Ground Level Enhancements of flux of the solar cosmic rays. The information on registration of single muons of the cosmic rays at the BUST was used for the analysis. During the period from April, 1981 to December, 2006, 36 GLE events were recorded by the world-wide network of neutron monitors. During 34 of them the muon data were registered at the BUST. These data were analyzed to search for short-term muon bursts.

The Baksan telescope is a parallelepiped with the basis 17 × 17 m$^2$ and with the height 11 m. All its sides are covered with scintillation detectors. Inside the telescope there are additional layers of scintillator. The effective area of the BUST is ≈200 m$^2$. The effective underground depth of its location is 850 m water equivalent (≈ 320 m of a rocky ground). Muons are registered at the BUST as trajectory events for which the directions of arrival ($\varphi, \vartheta$) are calculated. They were added up to angular distribution during every 15 minutes. The visible part of the sky was divided into 680 angular cells with the size 10° × 15° which partly overlapped. The counting rate of muons in each cell was analyzed within 3 hours during each GLE event (1 hour before the maximum of the X-ray flare and 2 hours after that). For comparison, the background intervals, when there were no GLE events and powerful solar flares, were analyzed by the same manner. In more detail the method of the analysis is described in [7, 8]. As a result, short-term muon bursts considerably exceeding the background were found at least in four GLE events: September 29, 1989; October 28, 2003; June 15, 1991 and October 12, 1981. They are enumerated in decreasing order of importance of the corresponding muon burst at the BUST.

The significant bursts are 1-2 hours delayed with respect to the maximum of the X-ray flare. The additional muon flux is narrow enough in space. Its angular diameter is about 10°-20°. Directions corresponding to the center of each burst are grouped in space in the interval 0°-60° of solar-ecliptic longitudes to the west from the Sun. It approximately corresponds to the direction of the force lines of the interplanetary magnetic field connecting the Sun with the Earth. The energy of the muons which caused bursts is ≥200 GeV. In this case the primary protons had the energy not lower than 500 GeV. It is approximately 100 times higher than the energy of registration at the neutron monitors which are usually used to study the SCR on the terrestrial surface. Both the nature of the described muon bursts and their possible relation to SCR are not quite clear so far. As was said above, the most energetic solar cosmic rays are generated on the Sun during powerful flares and processes accompanying them [1]. Registration of the particles accelerated in flares up to the greatest possible energy achievable on the Sun is one of the major tasks in the studies of the SCR acceleration processes [13]. The task solution will allow determining the upper limit and the form of the SCR spectrum at high energy. That is rather important for a lot of models of acceleration and emission of particles in the interplanetary space during powerful solar flares [1, 9-12]. As a rule, the GLE events have multistage development with complex interaction of various structures on surface and in crown of the Sun. The correct choice of acceleration model will allow obtaining the realistic scenario of development of the specified processes.

Later in the GLE event on July 14, 2000 similar muon bursts were recorded by the L3+C detector (CERN) [14] and the TEMP muon hodoscope (MEPhI) [15]. The differential energy spectrum in the event was very soft in comparison with other GLE events. At the description of spectrum by power function $J(E) \sim J_0 \cdot E^{-\gamma}$, the exponent γ was equal 6 on the signal increase phase and has risen up to 7-9 during the signal recession phase (the spectrum became even softer). In other words, the SCR spectrum in the event is described by the function abruptly falling with increase of energy of particles (for comparison, the GCR spectrum has γ ≈ 2.7). It is no wonder that muon bursts visibly exceeding the background was not recorded at the BUST in this GLE event, as energy of the particles registered by the BUST is essentially higher, than at L3+C and TEMP.

In the coming few years, when the LHC and ATLAS should start to operate, the increase in the solar activity is expected in the new 24th cycle. It increases probability of detection of the described muon bursts. Thus, there is a good opportunity to find out with the help of ATLAS whether there is causal relationship between the muon bursts and the solar cosmic rays.

## 3 Characteristics of ATLAS in comparison with the BUST

ATLAS is intended to verify the standard model and to search for a new physics at the Large Hadron Collider (LHC, CERN) [16]. In addition, it allows registration of muons of the cosmic rays (CR). ATLAS possesses an excellent muon system which allows searching for muon bursts during the GLE events similar to above ones. Moreover, ATLAS makes it possible not only to determine the initial direction of arrival of the CR muon but also to measure its momentum (energy) up to few TeV.

The design of the muon system of the ATLAS detector is shown in fig. 2. The external part of the muon system is a horizontal cylinder with the diameter 22 m and the length about 28 m [17]. Thus, the area of the detector for vertical flux of cosmic rays is 616 m$^2$. It is approximately 3 times larger than the area of the BUST. This allows the counting rate (and statistics) at ATLAS 3 times greater than at the BUST for the same flux of the CR muons. Moreover, ATLAS is approximately 4 times less deep underground (≈80 m) than the BUST (≈320 m). As a result, the minimal muon energy which is necessary to pass through the ground and to get in the detector will also be 4 times smaller, about 50 GeV. The energy of primary protons which generate muons in interactions in the atmosphere will also be approximately 4 times smaller. The integral spectrum of the cosmic rays is a steeply decreasing power function with the exponent α: $I(E) \sim I_0 \cdot E^{-\alpha}$. Therefore, the CR flux will increase many times with decreasing of muon energy necessary for registration. Namely, the muon flux registered at ATLAS will be approximately $4^\alpha$ times higher than at the BUST. Exponent α has different values for the galactic cosmic rays (GCR, α ≈ 1.7) and for SCR. In the latter case the exponent can vary from flare to flare and even during one GLE event from 2 to 6 and more. Therefore, the muon flux of GCR at the depth of ATLAS will be 10 times higher than at the depth of the BUST and 15 times and more for SCR (depending on the spectrum exponent).

Thus, apart from the general increase in the muon flux and statistics, the signal / background ratio should also be improved if the observed muon bursts are generated by the SCR. Considering the greater area of ATLAS, the number of registered GCR muons will be approximately 30 times larger than at the BUST, and no smaller than 45 times for SCR. While the muon bursts at the BUST are registered near the limit of statistical accuracy, ATLAS should register reliable excess of muons (short-term burst of intensity) during the GLE event.

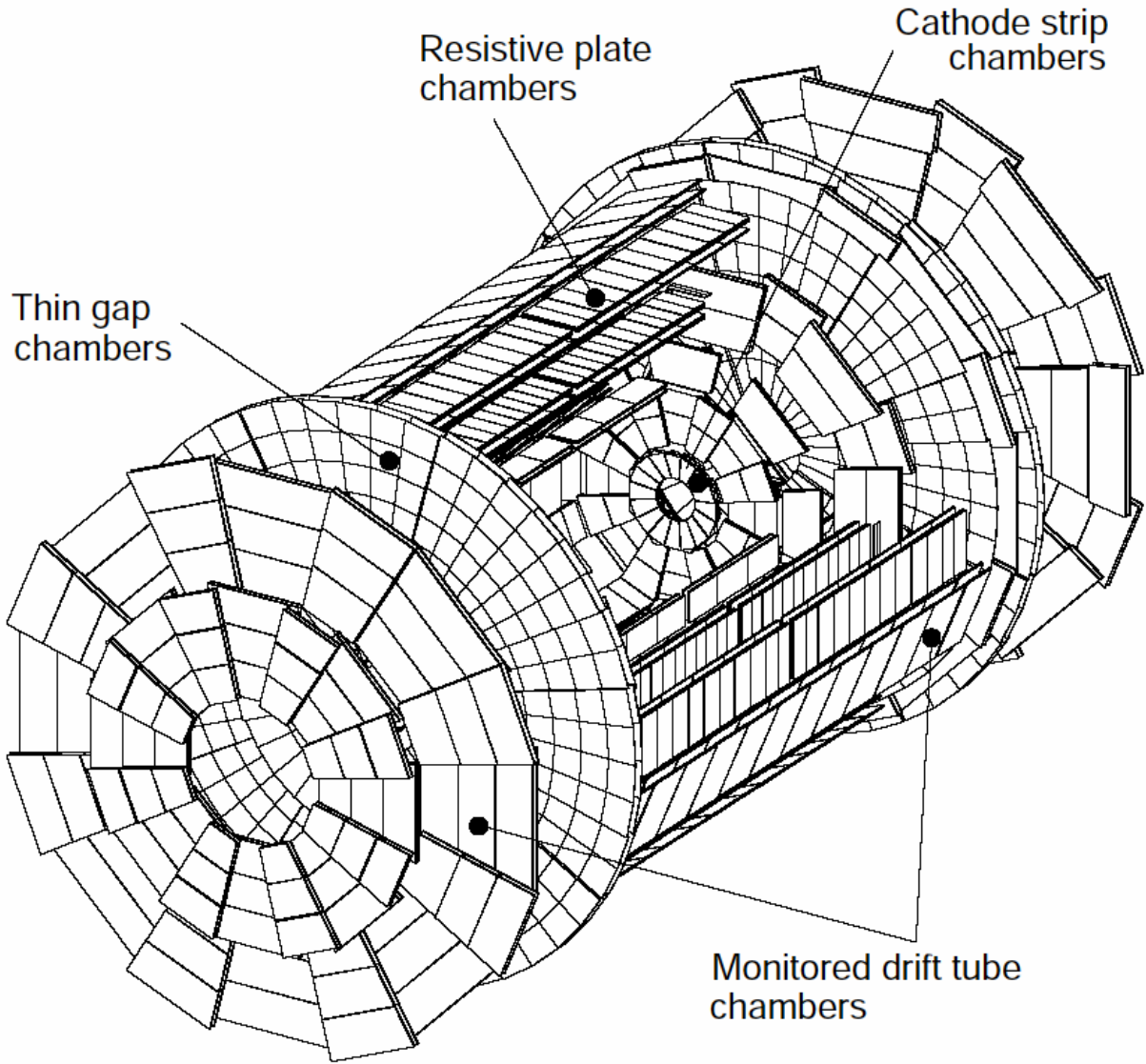

**Fig.2.** The muon system of the ATLAS detector [17].

The Ground Level Enhancements of SCR are rather rare events. Their average frequency is about 1 GLE event per year. Therefore it is obvious that continuous registration of the muon flux at ATLAS is necessary for searching for and studying of muon bursts during the GLE events. Intervals between sessions of exposure to the LHC beam are not enough for that. This is also confirmed by the L3+C collaboration experience [14]. In other words, it is necessary to register the cosmic ray muons by the ATLAS detector in parallel with acquisition of events from pp collisions and in the absence of the beam. However, it will practically lead to double the frequency of physical events which should be stored. The frequency of physical events from pp collisions is expected to be about 200 Hz after the 3-level trigger selection at high luminosity of the LHC. The frequency of the CR muon registration will be ~300 Hz but the size of each event is smaller.

A collision of bunches of the LHC beam (with frequency of 40 MHz) is usually considered as an event in ATLAS. There occur 20-25 pp collisions during each such event. These are mainly background events. The ATLAS registration system has no dead time, and the interval between bunches (≈25 ns) may be conditionally considered as the duration of the event. As a result, each CR muon will always coincide with one of the events. It will be mainly background events because the frequency of random coincidence of muons with physically interesting events which passed the trigger selection will be only ~0.003 Hz. In addition to searching for muon bursts, these background events can also be used for calibrations, alignment, and determination of stability, performance and effectiveness of subdetectors. It is very important in the long operation of the ATLAS detector. If necessary, these events can also be used for the background analysis. These events (minus muons themselves) will be a random sample of the background practically undistorted by the trigger selection since the CR muons are the casual Poisson flux.

The ATLAS trigger system and trigger menu already contain all necessary rules for selection of the CR muons. They have been developed by the muon group with the purpose of using CR for calibration and alignment of subdetectors in the absence of the LHC beam [18]. All characteristics of muon tracks which are necessary for the further analysis are calculated in the trigger of the second level ("TrigL2CosmicMuon" and "TrigL2CosmicMuonHypo"). The decision to record the event will be taken at the output of this trigger. The storage of an entire event (RAW data) is possible as is made now for calibration and alignment of subdetectors in the absence of the beam. As was pointed out above, it will lead to double the number of recorded events during the operation of the LHC. Yet, trade-off is also possible: the basic characteristics of each muon track in the ESD or AOD format or only the "muon vector" (time of event, momentum, angles $\theta$ and $\varphi$) will be stored. These data are enough to search for muon bursts during solar flares. In this case the size of the data will be appreciably smaller. However the events which were recorded in these formats will become practically useless for the purposes of calibration and alignment. In addition, if in the algorithm of the trigger analysis there are mistakes or discrepancies which are always probable, correction of the recorded events will be already impossible without full information.

This work is the proposal for using the cosmic ray muons which are registered by the ATLAS detector to obtain of the important CR physics information. Practically all tools and software necessary for this purpose are available. The ATLAS Collaboration might make the decision on continuous registration of the events containing the CR muons and about the format of the stored data. It will allow searching for short-term muon bursts during solar flares and studying other variations in the muon flux and also its spectrum and anisotropy.

## Conclusion

The ATLAS detector can be potentially used for studying of the solar cosmic rays (SCR), in particular for searching for short-term muon bursts, which were earlier found at the Baksan Underground Scintillation Telescope (BUST) during powerful solar flares. It would allow throwing light on possible relation of such bursts to the high energy SCR. It would also clarify the upper limit and the form of the energy spectrum of the SCR.

The possible signal should have greater amplitude both in absolute value of the muon counting rate and in the signal / background ratio in comparison with the BUST due to smaller thickness of ground above the ATLAS detector and due to its greater area. The energy of both the muons registered by ATLAS and the primary protons generating these muons in the atmosphere is on the average four times lower, than at the BUST. At the same time, the energy of primary protons at ATLAS will be tens of times higher than at the neutron monitors used usually for registration of the SCR.

The start-up of the LHC and ATLAS and the subsequent several years falls within the period of an increase in the solar activity. It raises the probability of finding the muon bursts from powerful flares on condition that continuous registration of the CR muons with recording of time and directions of arrival of muons at ATLAS will be provided. The solution of this task will allow determining the upper limit and the form of the SCR spectrum at high energy. That is rather important for a lot of models of acceleration and emission of particles in the interplanetary space during powerful solar flares [1, 9-12].